\documentclass[aps,prl,nobibnotes,twocolumn,superscriptaddress,floatfix]{revtex4-2}

\usepackage{amsmath}
\usepackage{amssymb}
\usepackage{bm}
\usepackage{braket}
\usepackage{color}
\usepackage{graphicx}
\usepackage[colorlinks=true,pdfstartview=FitV,linkcolor=blue,citecolor=blue,urlcolor=blue]{hyperref}

\def\ie{{\it i.e.},\ }

\makeatother

\begin{document}
	\title{Transverse Shifts Driven by Geometric Orbital Torque in Altermagnets}
	
	\author{Runze Li}
	\affiliation{Key Lab of advanced optoelectronic quantum architecture and measurement (MOE), Beijing Key Laboratory of Quantum Matter State Control and Ultra-Precision Measurement Technology, and School of Physics, Beijing Institute of Technology, Beijing 100081, China}
	
	\author{Yugui Yao}
	\affiliation{Key Lab of advanced optoelectronic quantum architecture and measurement (MOE), Beijing Key Laboratory of Quantum Matter State Control and Ultra-Precision Measurement Technology, and School of Physics, Beijing Institute of Technology, Beijing 100081, China}
	
	\author{Zhi-Ming Yu}
	\email{zhiming\_yu@bit.edu.cn}
	\affiliation{Key Lab of advanced optoelectronic quantum architecture and measurement (MOE), Beijing Key Laboratory of Quantum Matter State Control and Ultra-Precision Measurement Technology, and School of Physics, Beijing Institute of Technology, Beijing 100081, China}

	\begin{abstract}
	Anomalous transverse shifts in interface reflection are conventionally attributed to spin-orbit coupling or its pseudospin analogue. Here, we uncover a distinct geometric mechanism driven solely by an orbital torque arising from Fermi-surface anisotropy. During total reflection, an incident wave packet penetrates the anisotropic transmitted medium as an evanescent wave. There, the broken rotational symmetry about the interface normal exerts an orbital torque, and the accumulated angular impulse---the product of the torque and the dwell time---leads to a transverse spatial shift. Applying this mechanism to altermagnets without spin-orbit coupling, we show that the orbital torque is odd in spin. As a consequence,  the two spin channels are deflected to opposite sides of the incident plane, giving an intrinsic  spatial spin separation in altermagnets. The shift also inherits the angular symmetry of the altermagnetic spin splitting and is strongly enhanced near the onset of total reflection. Our work thus uncovers a new mechanism for transverse shifts, predicts a novel spin separation effect in altermagnets, and offers a direct probe of the altermagnetic order.
		
	\end{abstract}
	\maketitle
	
	{\emph{\textcolor{blue}{Introduction.--}}}
	Interface scattering is a ubiquitous process in optics, electronics, acoustics, and other fields. Under certain conditions, the scattered beam acquires an anomalous spatial shift relative to the incident beam, which can be decomposed into longitudinal~\cite{Goos1947,Renard1964,Beenakker2009,Haan2010,ZW2011} and transverse components~\cite{IF1955,IF1972}. Of the two, the transverse shift is highly nontrivial, as it drives the scattered beam out of the plane of incidence~\cite{Miller1972,sidejump2006S,KY2006,Hosten2008,bliokh2015,Jiang2015,Yang2015,Yliu2017,Yu2018,2LiuYing2018,shi2019,Ch2019,Yingliu2020,Do2022,RZ2024,YangMou2024,RZ2026}. Conventionally, this transverse shift originates from spin-orbit coupling (SOC) or pseudo-SOC, the latter stemming from a spin-like degree of freedom of the system. During scattering, a change in the (pseudo-)spin angular momentum is compensated by a corresponding change in the orbital angular momentum, leading to a transverse displacement~\cite{KY2006,Hosten2008,bliokh2015,Jiang2015,Yang2015,Yliu2017,Yu2018}. Consequently, transverse shifts in  systems without (pseudo-) SOC have received little attention.

	In this work, we identify a distinct mechanism for generating transverse shifts without SOC and pseudo-SOC. Consider an electron wave packet incident from an isotropic normal metal onto a medium whose band dispersion is anisotropic within the interface plane. Since the junction system does not possess continuous rotational symmetry about the interface normal, the corresponding component of the wave-packet orbital angular momentum (OAM) is not conserved during reflection. During total reflection, the incident wave enters the anisotropic transmitted  medium as an evanescent wave, remains there for a finite dwell time~\cite{Buttiker1983,Winful2003,Winful2009}, and is then reflected back.
	
	Altermagnets (AMs)~\cite{Tomas2022,2Tomas2022} provide a natural platform for studying this geometric mechanism. As a newly identified class of collinear magnets, their defining feature is the momentum-dependent spin splitting without SOC, despite the absence of net magnetization~\cite{Tomas2022,2Tomas2022,Qihang2022,Nicola2024,RunWu2024,Changyoung2024,FangChen2024,krempasky2024,takagi2025,Xunkai2025,zhou2025Nature,jiang2025metallic,Yichen2026,TianMu2026}. Depending on the crystal rotation symmetries connecting opposite spin sublattices, this spin splitting can exhibit $d$-, $g$-, or $i$-wave symmetry. A direct consequence of this unconventional magnetic order is the spin-dependent intrinsic anisotropy of the Fermi surface~\cite{Tomas2022,Tingli2024}. Thus, AMs without SOC provide a pristine environment to probe the existence of the transverse shift and its dependence on the Fermi-surface anisotropy.
	
	Here we demonstrate the transverse shift driven by geometric orbital torque in a minimal normal-metal/AM junction with neither spin-orbit coupling nor any pseudospin degree of freedom. The geometry here is distinct from quantum geometry of Bloch states (e.g.,Berry curvature), and arises solely from the intrinsic Fermi surface anisotropy. In this model, each spin channel is described by a single-band Hamiltonian with a completely trivial band structure. However, we find that even for such a model, a conventional quantum-scattering treatment yields a finite transverse shift. We trace this shift to the orbital torque and to the intrinsic anisotropy of the AM Fermi surface. During total reflection the incident beam penetrates the AM as an evanescent wave and accumulates an angular impulse that twists the reflected beam out of the plane of incidence. The orbital-torque formulation further shows that in AMs, the torque is odd in spin at every incident momentum, being equal in magnitude and opposite in sign for the two spin channels. Consequently, the transverse shift in the AM junction is accompanied by a robust spin separation, \ie the two spin components of an unpolarized incident beam are deflected to opposite sides of the plane of incidence during reflection. Our work thus uncovers a new mechanism of the transverse shift, shows that it gives rise to a spatial spin separation in AMs, and identifies this shift as a possible means to probe the symmetry of AMs.

\begin{figure}[t]
	\includegraphics[width=8.6cm]{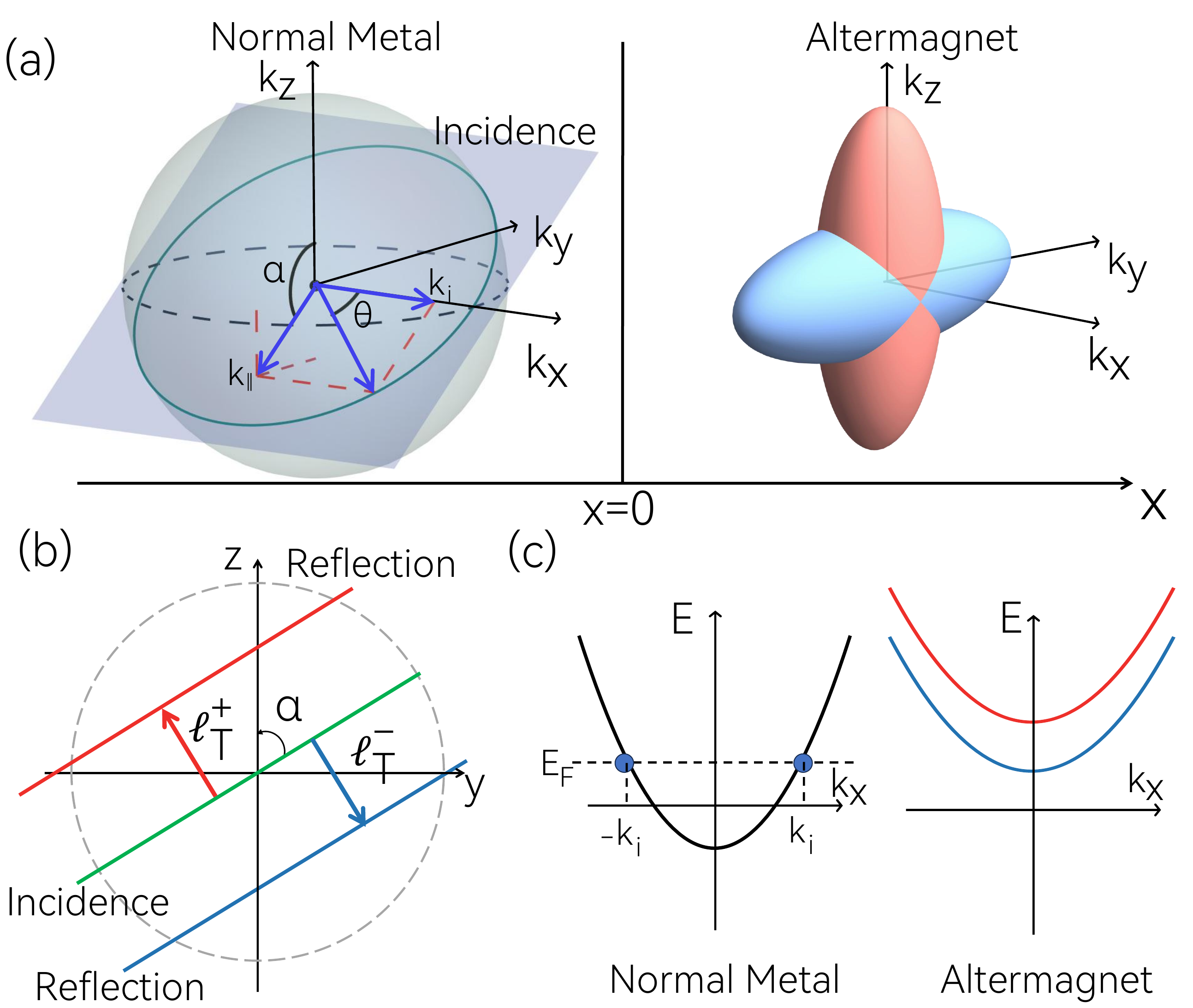} \caption{Geometry of a normal-metal/altermagnet junction. (a) Illustration of the spin-degenerate Fermi surface of the normal metal and spin-resolved Fermi surfaces of the altermagnet. The interface is  at $x = 0$. The blue plane is the incidence plane, spanned by the normal wave-vector component $k_i$ and the in-plane wave vector $k_\parallel$. (b) Illustration of the transverse shift $\ell_T^\pm$ projected on the $y$--$z$ interface plane. The two spin channels have opposite transverse shifts. (c) Schematic showing the spectrum of normal metal and altermagnet at a finite $k_{\parallel}$. Here, red and blue denote the spin-up and spin-down channels, respectively. \label{fig1}}
	\end{figure}

	{\emph{\textcolor{blue}{Effective model and transverse shift.--}}}
	We consider a flat interface at $x=0$ separating a normal metal ($x<0$) from an AM metal ($x>0$), as illustrated in Fig.~\ref{fig1}(a). Owing to the translational invariance of the junction in the $y$--$z$ plane, the interface wave-vector components $k_y$ and $k_z$ are conserved during scattering, while the normal component is affected.
	
	The wave vector of the incident beam $\boldsymbol{k}$ is characterized by two angles. The azimuthal angle $\alpha$ specifies the orientation of the incident plane relative to the $z$-axis $\alpha = \arg (k_z+ i k_y)$, and the incident angle within the incident plane is
	$\theta = \arg (k_i+ i k_{\parallel})$, where $k_{\parallel}=\sqrt{k_y^2+k_z^2}$ and $k_i$ is the wave vector component normal to the interface.
	
	The normal metal occupying the $x<0$ region is described by a spin-degenerate three-dimensional (3D) electron gas (we set $\hbar=1$),
	\begin{equation}\label{EGas}
		H_{L} = \left(\frac{\boldsymbol{k}^{2}}{2m} - U_{L}\right)\sigma_{0},
	\end{equation}
	where $m$ is the effective mass, $U_L$ is the potential, and $\sigma_0$ is the identity matrix acting on spin space. In the absence of SOC, the two spin channels are decoupled, and each contains an incident state with $k_x=k_i$ and a corresponding reflected state with $k_x=-k_i$, where $k_i=\sqrt{2m(E_F+U_L)-k_{\parallel}^2}$.
	
	For the right-hand region ($x>0$), we consider a minimal AM Hamiltonian without SOC,
	\begin{equation} \label{genmodel2}
		H_{R} = (a\boldsymbol{k}^{2}-U_{R})\sigma_{0} + b k_{\parallel}^n\cos[n(\alpha+\beta)]\sigma_{z},
	\end{equation}
	where $n$ is an even integer, $\sigma_{z}$ is the Pauli matrix, and $\beta$  denotes the phase offset of the spin degenerate points. The orders $n=2,4,6$ represent the $d$-, $g$-, and $i$-wave AMs, respectively. Here, we retain the leading planar spin splitting and choose the interface normal $\hat{x}$ along the high-symmetry axis. Since spin is a good quantum number in both regions, the junction Hamiltonian decouples into two independent spin channels (labeled $\pm$):
	\begin{equation} \label{junc}
		{\cal H}=
		\begin{cases}
			H_{L}^{\pm} = \dfrac{\boldsymbol{k}^{2}}{2m} - U_{L}, & x < 0,\\[8pt]
			H_{R}^{\pm} = (a\boldsymbol{k}^{2}-U_{R}) \pm b k_{\parallel}^n\cos[n(\alpha+\beta)], & x > 0,
		\end{cases}
	\end{equation}
	In the AM region, the two spin-resolved Fermi surfaces are related by a $\pi/n$ rotation about the interface normal, \ie $H_R^{+}(\alpha)=H_R^{-}(\alpha-\pi/n)$.
	
	Under total reflection, the transmitted waves are evanescent, with a purely imaginary wave vector $k_x = i k_s^\pm$, and $k_s^\pm	=\sqrt{a  k_{\parallel}^{2}\pm b k_{\parallel}^{n}\cos\!\left[n(\alpha+\beta)\right]-(E_F+U_R)}/\sqrt{a}.$ $k_s^\pm$ determines the penetration depth and encodes the spin-resolved Fermi-surface anisotropy. The scattering state of the junction [Eq.~\eqref{junc}] therefore takes the form
	\begin{equation} \label{eq:psi}
	\Psi^\pm(\boldsymbol{r}) =
		\begin{cases}
		\left( e^{i k_i x} + r^\pm e^{-i k_i x} \right) e^{i \boldsymbol{k}_{\parallel} \cdot \boldsymbol{r}_{\parallel}}, & x < 0,\\
		t^\pm e^{-k_s^\pm x} e^{i \boldsymbol{k}_{\parallel} \cdot \boldsymbol{r}_{\parallel}}, & x > 0,
	\end{cases}
	\end{equation}
	where $r^\pm$ and $t^\pm$ are the reflection and transmission coefficients, respectively. At the interface $x=0$ we impose the BenDaniel--Duke boundary conditions \cite{BenDaniel1966}, namely the continuity of the wave function and of the effective-mass-weighted derivative $(1/m^{*})\partial_x\Psi^\pm$: $\Psi^\pm(0^-)=\Psi^\pm(0^+)$ and  $m^{-1}\,\partial_x\Psi^\pm(0^-)=2a\,\partial_x\Psi^\pm(0^+)$, where $m^{*}=m$ for $x<0$ and $m^{*}=1/(2a)$ for $x>0$ with $a>0$.
	Since the probability-current density is
	$j_x=(1/m^{*})\,\mathrm{Im}[\Psi^{\pm *}\,\partial_x\Psi^\pm]$, these conditions
	guarantee its conservation across the interface. Solving the boundary conditions gives
	\begin{equation} \label{eq:rt}
	r^\pm = \frac{k_i - i (2 a m k_s^\pm)}{k_i + i (2 a m k_s^\pm)}, \qquad
	t^\pm = \frac{2 k_i}{k_i + i (2 a m k_s^\pm)}.
	\end{equation}
	
	Within the standard approach~\cite{Jiang2015,Yang2015,Yu2018}, the transverse shift, illustrated in Fig.~\ref{fig1}(b), is expressed as
	\begin{equation}
	\ell_T^\pm = \ell_z^\pm \sin\alpha-\ell_y^\pm \cos\alpha
	= \frac{1}{k_{\parallel}}\frac{\partial \phi^\pm}{\partial \alpha},
	\end{equation}
	where $\phi^\pm$ is the phase of the reflection coefficient $r^{\pm}=e^{i\phi^\pm}$ in the total-reflection regime, given by $\phi^\pm = -2 \arctan\left(2 a m k_s^\pm/k_i \right)$.
	Then, the transverse shift is obtained as
	\begin{equation}\label{TransverseShift}
	\ell_T^\pm = \pm \frac{2mnbk_ik_{\parallel}^{n-1}}{k_s^\pm (k_i^2 + 4a^2m^2 (k_s^\pm)^2)} \sin[n(\alpha+\beta)].
	\end{equation}
	This result shows that the transverse shift can emerge in the system without (pseudo-)SOC or nontrivial band topology, in contrast to previously identified mechanisms~\cite{Jiang2015,Yang2015,Yu2018}. As shown below, the shift is solely driven by a  geometric orbital torque accumulated during the evanescent dwell time.

	{\emph{\textcolor{blue}{Geometric orbital torque.--}}}
	We now identify the physical origin of the shift. A transverse shift $\ell_T$ is equivalent to a change of the OAM about the interface normal ($\Delta L_x$), given as
	\begin{equation}\label{ellT2}
		\ell_T=-\Delta L_x/k_{\parallel}.
	\end{equation}
	This means that the transverse shift is closely related to the dynamics of $L_x$. Since spin is conserved in the absence of SOC for collinear magnets, $L_x$ can change only through a torque on the orbital degree of freedom, i.e., the orbital torque.
	
	For the junction model studied here, $L_x$ is a conserved quantity in the incident region.  However, the intrinsic in-plane anisotropy of the AM [Eq.~\eqref{genmodel2}] breaks the continuous rotational symmetry about the interface normal, rendering $[\hat{H},\hat{L}_x]\neq 0$. As a consequence, the AM exerts an orbital torque on the evanescent wave packet during its dwell time, delivering the accumulated change of the OAM to the reflected wave and thus leading to a finite transverse shift.
	
	To formalize this, we express the OAM operator as
	\begin{equation}
	\hat{L}_x = (\hat{\boldsymbol{r}}\times\hat{\boldsymbol{p}})_x
	= i\left(k_z\frac{\partial}{\partial k_y}
	- k_y\frac{\partial}{\partial k_z}\right)
	= i\frac{\partial}{\partial\alpha}.
	\end{equation}
	The time evolution of $\hat{L}_x$ is governed by the Heisenberg equation of
	motion,
	\begin{equation}
	\frac{d\hat{L}_x}{dt}= i[\hat{H},\hat{L}_x]\equiv\hat{\mathcal{T}}_x,
	\end{equation}
	which defines the orbital torque operator.  From Eq. (\ref{junc}), one has
	$\hat{\mathcal{T}}_x=-\Theta(x)\,nbk_{\parallel}^{n}\sin\!\left[n(\alpha+\beta)\right]\sigma_z$,
	with $\Theta(x)$ the step function. Projecting this operator onto the two spin
	eigenstates yields the spin-resolved torque
	\begin{equation}\label{OAMtorque}
		\mathcal{T}_x^\pm = \mp nbk_{\parallel}^{n}\sin[n(\alpha+\beta)],
	\end{equation}
	which is another key result of this work.
	
	\begin{figure}[t]
	\includegraphics[width=8.6cm]{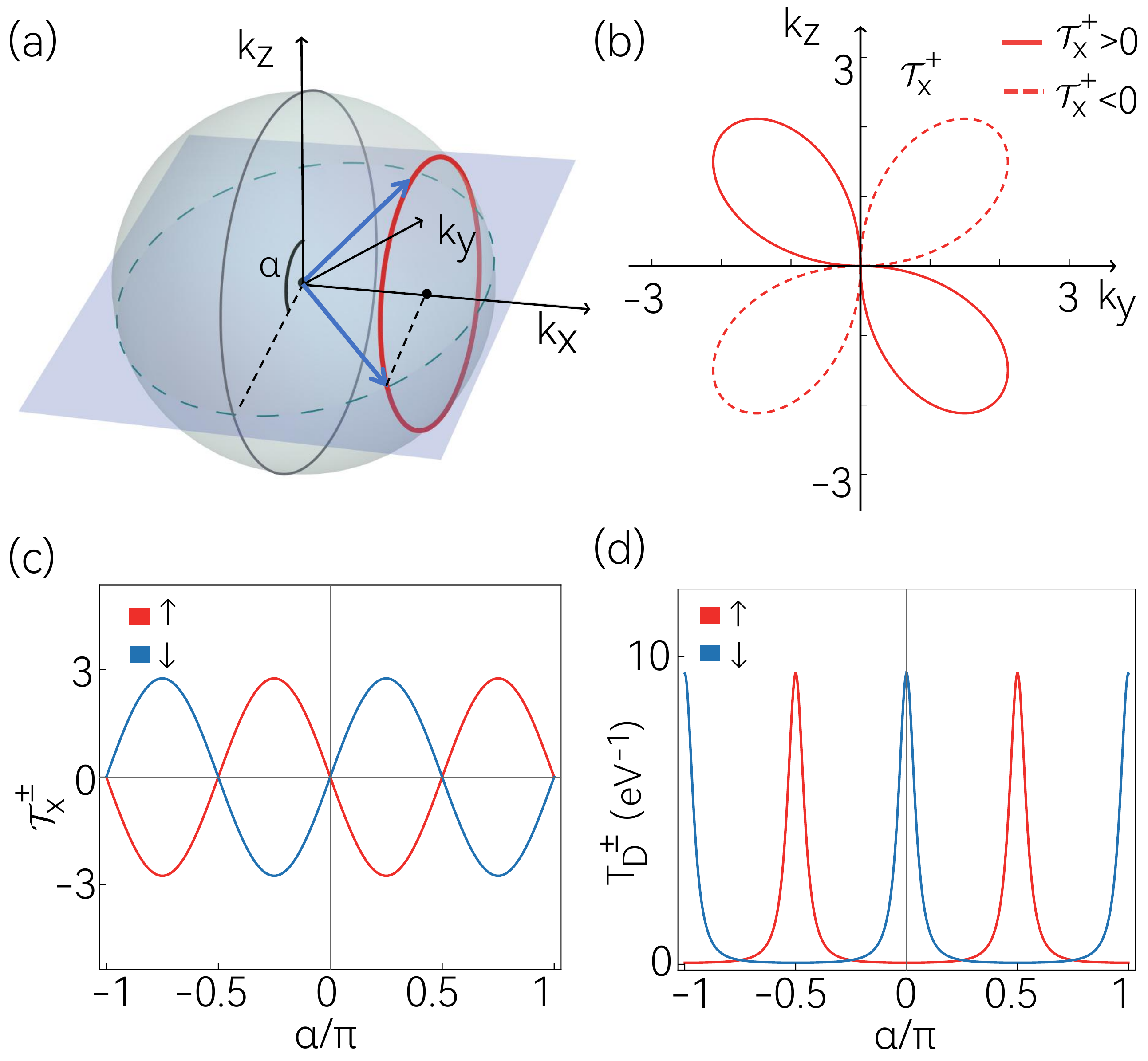} \caption{(a) The incident beam with fixed $k_{\parallel}$ but variable $\alpha$ traces a circle on the Fermi surface of the normal metal, shown as the red circle. Here $k_{\parallel}=0.7R$, where $R=\sqrt{2m(E_F+U_L)}$ is the Fermi radius. The solid circle on the sphere marks the boundary of accessible interface wave vectors in the normal metal. (b-d) The  torque and the dwell times as functions of $\alpha$ at fixed $k_{\parallel}=0.7R$. (b) Polar representation of  the  torque $\mathcal{T}_x^{+}$ for spin-up, showing a $d$-wave feature. (c) and (d) show the spin-resolved torques $\mathcal{T}_x^{\pm}$ and dwell times $T_D^\pm$, respectively. We set $n=2$, $a=2\,\mathrm{eV\,nm^2}$, $b=1.8\,\mathrm{eV\,nm^2}$, $m=0.1m_e$, $E_F=0.1\,\mathrm{eV}$, $U_L=0.5\,\mathrm{eV}$, $U_R=0$, and $\beta=0$.\label{fig2}}
	\end{figure}

	During total reflection, the incident wave penetrates the AM as an evanescent wave for a finite dwell time before being totally reflected, and throughout this process it experiences the geometric torque only within the transmitted region. The total accumulated OAM impulse is therefore the product of the torque and the dwell time,
	\begin{equation}
		\Delta L_x^\pm = \mathcal{T}_x^\pm\, T_D^\pm,
	\end{equation}
	where $T_D^\pm$ is the dwell time of the evanescent wave in the AM, defined as~\cite{Buttiker1983,Winful2003} $T_D^\pm = \int_{x>0}|\Psi_R^\pm(x)|^2\,dx/j_{\mathrm{in}}$, with
	$\Psi_R^\pm(x)=t^\pm e^{-k_s^\pm x}$ the  transmitted wave function in the AM, and $j_{\mathrm{in}}=k_i/m$ the incident probability-current density normal to the interface. A direct calculation yields
	\begin{equation}
	T_D^\pm = \frac{2m k_i}{k_s^\pm\,(k_i^2 + 4a^2m^2 (k_s^\pm)^2)},
	\end{equation}
	which are positive definite and related by a $\pi/n$ rotation about the interface normal, $T_D^+(\alpha)=T_D^-(\alpha-\pi/n)$.
	The accumulated OAM impulse is entirely converted into a transverse spatial shift via Eq. (\ref{ellT2}).
	One can check that the shift $\ell_T^\pm$  obtained from the orbital-torque approach coincides with that of the conventional method \eqref{TransverseShift}, confirming the
	consistency of the two approaches.
	
\begin{figure}[t]
	\includegraphics[width=8.6cm]{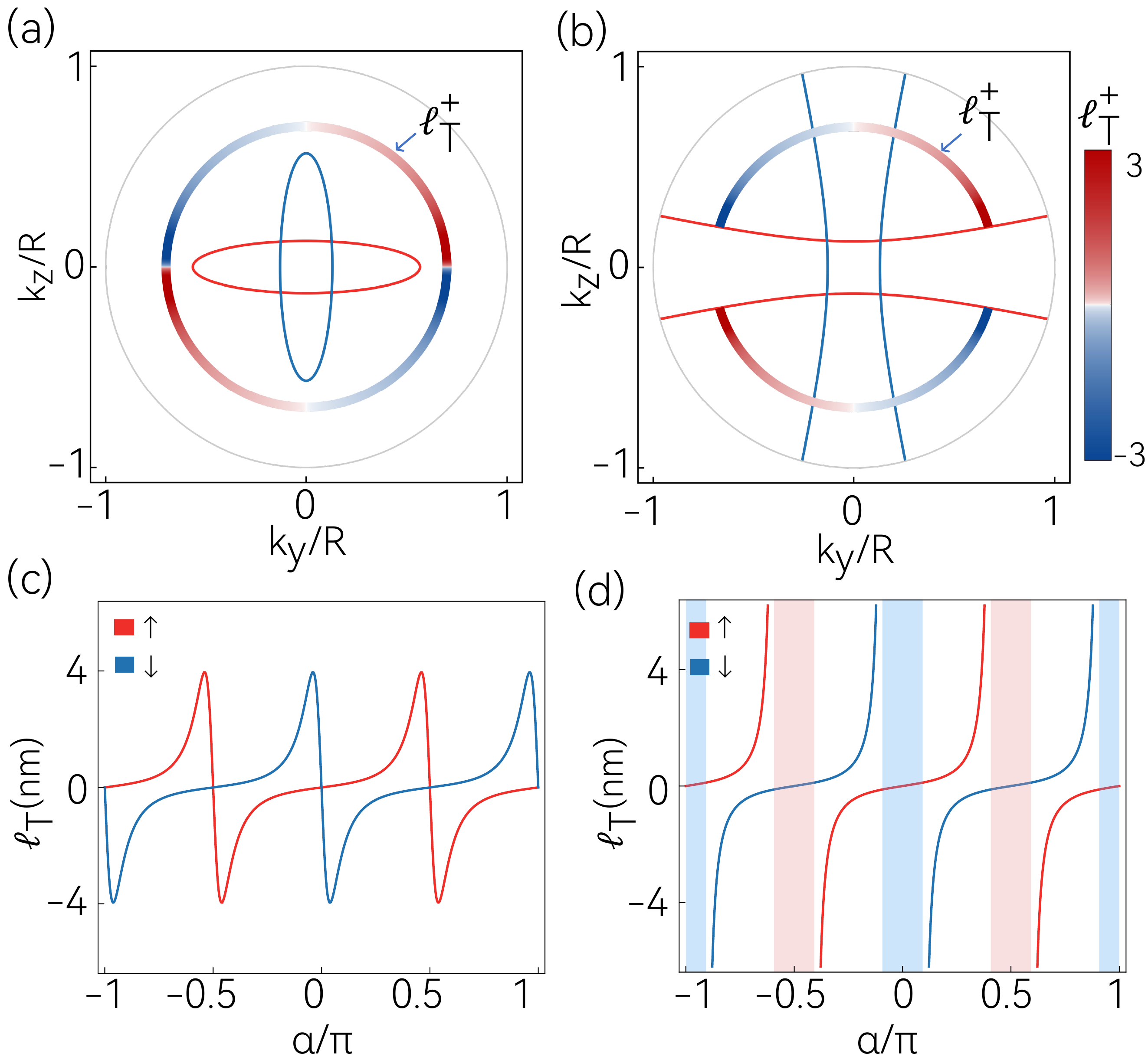} \caption{(a,b) Numerical results of the transverse shift $\ell_T^+$ for spin-up along the red circle in Fig. \ref{fig2}(a). The values of $\ell_T^+$  are indicated by the color. The outer gray solid line represents the boundary of the projected Fermi surface of the normal metal, corresponding to the gray solid circle in  Fig. \ref{fig2}(a). The inner red (blue) curves are the boundary of the projected spin-up (spin-down) Fermi surface of the altermagnets. (c,d) Numerical results of the transverse shifts $\ell_T^+$ and $\ell_T^-$. The parameters are $n=2$, $m=0.1m_e$, $E_F=0.1\,\mathrm{eV}$, $U_L=0.5\,\mathrm{eV}$, $U_R=0$, and $\beta=0$. For (a,c), we set $a=2\,\mathrm{eV\,nm^2}$, $b=1.8\,\mathrm{eV\,nm^2}$; for (b,d), we set $a=1.8\,\mathrm{eV\,nm^2}$, $b=2\,\mathrm{eV\,nm^2}$.\label{fig3}}
	\end{figure}

	Several important characteristics follow directly from the orbital torque induced transverse shift. (i) The transverse shift can solely originate from the anisotropy of the systems.
	Thus, due to intrinsic altermagnetic anisotropy, transverse shift can naturally appear in the interface between normal metals and AMs.
	(ii) From Eq. (\ref{OAMtorque}), one observes that at fixed incident momentum, the two spin channels in AMs experience equal and opposite torques, as shown in Fig.~\ref{fig2}(c).
	Since the dwell time is always positive, the two spin channels must acquire opposite transverse shifts. However, the shifts for the  two spin channels are generically asymmetric in magnitude, due to unequal dwell times [see Fig.~\ref{fig2}(d)].
	(iii) The orbital torque $\mathcal{T}_x^\pm$ and the shift $\ell_T^{\pm}$ inherit the  symmetry of the junction Hamiltonian. For our model, the factor $\sin[n(\alpha+\beta)]$ in $\mathcal{T}_x^\pm$ and then $\ell_T^{\pm}$ enforces $2n$ symmetry-protected zero-points at $\alpha=-\beta+m\pi/n$, which mark the principal-axis directions where the spin splitting is maximal [see Figs.~\ref{fig2}(c) and~\ref{fig3}(c)]. Thus, the shift provides a possible means to probe the symmetry of AMs.
	(iv) The shift is strongly enhanced near the onset of total reflection, where $k_s^\pm\to0$ prolongs the dwell time and thereby the OAM accumulation.

	Notably, the spin-opposite torque is not an artifact of the minimal model in Eq.~\eqref{genmodel2}, but is a general feature of AMs whenever the altermagnetic spin splitting dominates the Fermi-surface anisotropy. In the ideal case, the spin-independent background $H_0$ is isotropic, making $\partial_\alpha H_0=0$ and $\mathcal{T}_x^+=-\mathcal{T}_x^-$. Here, the spin-opposite torque arises solely from the altermagnetic-splitting term $H_z$. In the generic case, by contrast, $H_0$ is also anisotropic and contributes a finite orbital torque to both spin channels. However, as long as $|\partial_\alpha H_z|>|\partial_\alpha H_0|$, the total orbital torques on the two spins remain opposite in sign, giving rise to opposite-sign transverse shifts.
	Consequently,  interface reflection  provides a natural and robust route to generate a transverse spin current in AMs, with the two spin channels deflected in opposite directions and each deflected component remaining fully spin-polarized.

	{\emph{\textcolor{blue}{Transverse shift in a $d$-wave AM.--}}}
	To illustrate this mechanism concretely, we consider a $d$-wave AM ($n=2$) with nodal offset $\beta=0$.
	In this case $H_R^{\pm}$ reduces to an anisotropic parabolic band with the effective-mass tensor $[m_x,\,m_y^{\pm},\,m_z^{\pm}]=[1/(2a),\,1/(2a\mp 2b),\,1/(2a\pm 2b)]$. Then the transverse shift
	\eqref{TransverseShift} can  be simplified  as
	\begin{equation}\label{ellT3}
		\ell_T^\pm = \Pi^\pm\, \frac{\epsilon^\pm - 1}{k_s^\pm}\, \sin 2\alpha,
	\end{equation}
	where $\Pi^\pm = m m_x^2 k_{\parallel} k_i / \{m_y^\pm\,[m_x^2 k_i^2 + m^2 (k_s^\pm)^2]\}$. The dimensionless parameter $\epsilon^\pm = (a \pm b)/(a \mp b)$ characterizes the intrinsic in-plane anisotropy of the spin-resolved Fermi surface.
	
	For $|a|>|b|$, the constant-energy contours of the two spin channels are mutually perpendicular ellipses [Fig.~\ref{fig3}(a)], with $\epsilon^+\epsilon^-=1$ reflecting the mutual $\pi/2$ rotation. Thus, the geometric factor $(\epsilon^\pm - 1)$ inherently possesses opposite signs for the two spin channels. This again guarantees that the transverse shifts for the two spin channels are always opposite, as manifested in Eq. (\ref{ellT3}).
	The result of $\ell_T^\pm$ is plotted in Fig.~\ref{fig3}(c), which explicitly demonstrates the following properties. (i) The difference of the shift $\Delta\ell_T=\ell_T^+-\ell_T^-$ inherits the symmetry of the altermagnetic spin splitting, but is rotated by $\pi/4$ relative to it. Both $\ell_T^\pm$ and $\Delta\ell_T$ vanish along the directions $\alpha=m\pi/2$, where the spin splitting is maximal. Meanwhile, the spin splitting vanishes for
	\mbox{$\alpha=\pi/4+m\pi/2$}, where
	\mbox{$\mathcal{T}_x^\pm$} takes maximal values.
	(ii) The dwell times $T_D^+$ and $T_D^-$ are maximal at $\alpha=\pi/2+m\pi$ and $\alpha=m\pi$, respectively, at which $\mathcal{T}_x^\pm=0$. For small deviations from these angles, the torque magnitude grows as the dwell time decreases, yielding maximal transverse shifts of about 4 nm for suitable model parameters, as shown in Fig.~\ref{fig3}(c).
	
	In contrast, for $|a|<|b|$, the constant-energy contours open into mutually perpendicular hyperbolas, as shown in Fig.~\ref{fig3}(b).
	Here $\epsilon^\pm<0$ and  $|\epsilon^\pm|=\tan^2\alpha_{\mathrm{as}}^\pm$ with $\alpha_{\mathrm{as}}^\pm$ the azimuthal angle of the asymptotes, which obey $k_y=\pm\sqrt{|\epsilon^\pm|}\,k_z$.
	As $\alpha$ sweeps over $2\pi$ at fixed $k_\parallel$, each spin channel alternates between two propagating and two total-reflection sectors, and the shift vanishes in the propagating sectors \footnote{There, the transmitted longitudinal wave number $q^\pm=ik_s^{\pm}$ is real, so $r^\pm=(k_i-2amq^\pm)/(k_i+2amq^\pm)$ is real and $\ell_T^\pm\propto\partial_\alpha\arg r^\pm=0$.}, as shown in  Fig.~\ref{fig3}(d). In the large-$k_\parallel$ limit, the boundaries of the suppressed zones, where $\ell_T^{\pm}=0$, approach the asymptotic direction $\alpha_{\mathrm{as}}^\pm$. Thus, the locations and widths of the suppressed zones determine the geometry of the open Fermi surface, including the crystalline orientation $\beta$ and the hyperbolic anisotropy $\epsilon^\pm$.

	{\emph{\textcolor{blue}{Discussion.--}}}
	We have identified a purely geometric mechanism for anomalous transverse shifts in interface reflection. AMs provide a natural platform for realizing this mechanism, since the orbital torque therein is odd in spin and drives a deterministic spin separation under total reflection. The proposed shift also offers a direct probe of the altermagnetic order.
	
	Besides the even-wave magnets, odd-wave magnets such as $p$-wave magnets have recently been proposed~\cite{hellenes2023,Brekke2024,FloquetOddParity2026}, which, in contrast to their even-wave counterparts, are typically realized in noncollinear magnets~\cite{song2025,yamada2025}. Since odd-wave magnets possess both an intrinsic anisotropic Fermi surface and an effective SOC, the transverse shift therein may combine the orbital-torque and (pseudo-)SOC mechanisms, yielding rich and distinct shift behavior.
	
	Experimentally, the effect can be detected by injecting a spin-degenerate electron beam from a normal metal whose Fermi surface is significantly larger than that of the AM, which allows total reflection over a broad range of incident angles. The spatial separation of the opposite spin channels after reflection drives a transverse spin current, which can be converted into
	a measurable voltage via the inverse spin Hall effect (ISHE) in a Pt electrode~\cite{ES,ES2}.
	
	Finally, the mechanism uncovered here is not restricted to electronic systems. Altermagnetism has recently been emulated in classical wave platforms, including photonic and phononic systems~\cite{KIM2026,chen2026altermagnetic,xia2026time}, where the role of electron spin is played by a pseudo-spin degree of freedom. Because the transverse shift we derive relies only on the generic geometry of an anisotropic pseudo-spin-split band, it can be applied  directly to these classical analogs, and may lead to a pseudo-spin-opposite transverse shift,  offering a generic route to manipulating pseudo-spin.
	
	{\emph{\textcolor{blue}{Acknowledgments.--}}}
	This work is supported by the National Key R\&D Program of China (Grant 
	No. 2025YFA1411201), and  the National Natural Science Foundation of China (Grant Nos. 12474040, W2511003, and 12321004).
	
	\bibliography{cite}
	\bibliographystyle{apsrev4-2}
	
\end{document}